# Advanced MST3 Encryption scheme based on generalized Suzuki 2-groups

Gennady Khalimov[1][0000-0002-2054-9186], Yevgen Kotukh [2][0000-0003-4997-620X]

[1]Kharkiv National University of Radioelectronics, Kharkiv, 61166, Ukraine hennadii.khalimov@nure.ua
[2]Yevhenii Bereznyak Military Academy, Kyiv, Ukraine yevgenkotukh@gmail.com

**Abstract.** This article presents a method for enhancing the encryption algorithm in the MST3 cryptosystem for generalized Suzuki 2-groups. The conventional MST cryptosystem based on Suzuki groups utilizes logarithmic signatures (LS) restricted to the center of the group, resulting in an expansive array of logarithmic signatures. We propose an encryption scheme based on multi-parameter non-commutative groups, specifically selecting multi-parameter generalized Suzuki 2-groups as the group construction framework. In our approach, the logarithmic signature extends across the entire group, with cipher security dependent on the group order. This design enables the development of encryption optimized for implementation efficiency—determined by logarithmic signature size—while maintaining robust security through appropriate key sizes and the finite field of group representation. The primary innovation in our encryption implementation lies in the sequential de-encapsulation of keys from ciphertext using logarithmic signatures and associated keys. The security evaluation of the cipher relies on attack complexity analysis, which is quantified through comprehensive key enumeration methodologies.

## INTRODUCTION

Practical advancements in quantum computing development and the consequent tangible threat of quantum algorithms compromising modern cryptosystems have intensified research into public-key cryptosystems based on alternative difficult-to-solve problems beyond the classical approaches of integer factorization and discrete logarithms. Numerous cryptosystems exploring various complex group theory problems were proposed and examined during the 2000s [1-4].

The word problem represents one such challenging task [5]. Magliveras [6] proposed one of the first implementations in this paradigm, based on logarithmic signatures for finite permutation groups, which was subsequently extended by Lempken et al. for asymmetric cryptography with random covers [7]. This innovative approach expanded the hard-to-solve word problem to a diverse range of groups. The initial implementation of such a cryptosystem, termed MST3, was designed for the Suzuki group, with several enhancements documented in [8-9].

Subsequently, Svaba et al. [10] conducted a comprehensive security analysis of MST cryptography and introduced security improvements utilizing homomorphic transformation, an approach known as the eMST3 cryptosystem. T. van Trung [11] employed an alternative strategy to enhance overall security by proposing a method for strong aperiodic logarithmic signatures design, specifically for abelian p-groups.

MST cryptosystems based on multiparameter non-commutative groups were introduced in [12-15]. These cryptosystems fundamentally focus on overhead optimization, achievable through key length reduction and encryption/decryption algorithm acceleration. Research demonstrated that cryptosystems with logarithmic signatures spanning the entire group could be constructed on large-order groups over small finite fields. The generalized Suzuki 2-group, being a multiparameter group, possesses a larger group order over a fixed finite field of definition, providing an advantage compared to the classical Suzuki group. The initial cryptosystem implementation for the generalized Suzuki 2-group presented in [13] neither provided encryption for the entire Suzuki 2-group nor protection against brute force attacks with sequential key recovery.

This article presents an improved MST-based encryption implementation for Suzuki multiparameter 2-groups. We outline the basic encryption scheme, our contribution to its enhancement, and a preliminary security analysis. We acknowledge this as a broader research domain and aim to provide a comprehensive security analysis of the proposed solution in our future research.

## BASIC APPROACH

The MST cryptosystem is constructed upon the following fundamental principles [7]. Consider $G$ as an ultimate nonabelian group. The group has a nontrivial center $Z$, so that $G$ does not decompose over $Z$. Assume that $Z$ is large enough, so that the search over $Z$ is computational impracticable. If

$\alpha = [A_1,...,A_s]$ is a logarithmic signature (cover) then each element $g \in G$ can be expressed uniquely as a product of the form $g = a_1 \cdot a_2 \cdots a_s$, for $a_i \in A_i$. $\alpha = [A_1,...,A_s]$ is called tame (factorizable) if it can be factorized in a polynomial in the width $w$ of $G$.

The cryptographic hypothesis, which is the basis for the cryptosystem, is that if $\alpha = [A_1, A_2,..., A_s] := (a_{i,j})$ – accidental cover for a "large" mattress $s$ at $G$, then search for the layout $g = a_{1j_1} a_{2j_2} \cdots a_{sj_s}$ is computationally impracticable for any element $g \in G$ relatively to $\alpha$.

The generalizations of Suzuki 2-groups are defined over a finite field $F_q$, $q = 2^n$, $n > 0$ for a positive integer $l$ and $a_1, a_2,..., a_l \in F$ for some automorphism $\theta$ of $F$ as [16].

$$A_l(n,\theta) = \{S(a_1, a_2,..., a_l) \mid a_i \in F_q\}.$$

For each element of $A_l(n,\theta)$ can be expressed uniquely and it follows that $|A_l(n,\theta)| = 2^{nl}$ and $A_l(n,\theta)$ defines a group of order $2^{nl}$. If $l = 2$, this group is isomorphic to a Suzuki 2-group $A(n,\theta)$.

The group operation is defined as a product

$$S(a_1,a_2,...,a_l)S(b_1,b_2,...,b_l) = S(a_1 + b_1, a_2 + (a_1\theta)b_1 + b_2, a_3 + (a_2\theta)b_1 + (a_1\theta^2)b_2 + b_3,..., a_l + (a_{l-1}\theta)b_1 + ... + (a_1\theta^{l-1})b_{l-1} + b_l).$$

The identity element is $S(0_1, 0,..., 0)$.

The inverse element is

$$S(a_1, a_2, a_3,..., a_l)^{-1} = S(a_1, a_2 + a_1\theta a_1,$$
$$a_3 + a_2\theta a_1 + a_1\theta^2(a_2 + a_1\theta a_1),..., a_l + a_{l-1}\theta a_1 + ...).$$

The group $G$ is a nonabelian group and has a nontrivial center

$$Z(G) = \{S(0,0,...,c) \mid c \in F_q\}.$$

Assume that $\theta$ is the Frobenius automorphism of $F, \theta: x \to x^2$. For the fixed finite field, the order of group $A_l(n,\theta)$ is higher than the regular Suzuki 2 - group. In [13] authors gave enough presentation of basic encryption scheme based on the generalized Suzuki group.

Let $A_l(n,\theta) = \{S(a_1, a_2, ..., a_l) \mid a_i \in F_q\}$ be a large group, $q = 2^n$ with the center $Z$.

Step 1. Choose a factorizable logarithmic signatures $\beta_k = [B_{1(k)}, ..., B_{s(k)}] = (b_{ij})_k = S(0, ..., b_{ij(l/2+k)}, 0, ...)$ of type $(r_{1(k)}, ..., r_{s(k)})$, $i = \overline{1,s}$, $j = \overline{1, r_{i(k)}}$, $b_{ij(l/2+k)} \in F_q$, $k = \overline{1, l/2}$.

The tame logarithmic signature is defined as a bijective and factorizable map of $\beta_k(R)$.

Step 2. Select a random cover $\alpha_k = [A_{1(k)}, ..., A_{s(k)}] = S(0, ..., a^{(1)}_{ij(k)}, 0, ..., a^{(2)}_{ij(l/2+k)}, 0, ...)$ of the same type as $\beta$, where $a_{ij} \in A_l(n, \theta)$, $a^{(1)}_{ij(k)}, a^{(2)}_{ij(k)} \in F_q \setminus \{0\}$, $i = \overline{1,s}$, $j = \overline{1, r_{i(k)}}$, $k = \overline{1, l/2}$.

Step 3. Choose $t_{0(k)}, ..., t_{s(k)} \in A_l(n, \theta) \setminus Z$, $t_{i(k)} = S(t_{i1(k)}, ..., t_{il(k)})$, $t_{ij(k)} \in F^\times$, $i = \overline{0, s}$, $j = \overline{1, l}$, $k = \overline{1, l/2}$. Let's $t_{s(v)} = t_{0(v+1)}$, $v = \overline{1, l/2 - 1}$.

Step 4. Construct a homomorphism $f$ defined by
$$f(S(a_1, ..., a_l)) = S(0, ..., 0, a'_{l/2+1} = a_1, ..., a'_l = a_{l/2}).$$

Step 5. Calculating $\gamma_k = [h_{1(k)}, ..., h_{s(k)}] = t^{-1}_{(i-1)(k)} f((a_{ij})_k)(b_{ij})_k t_{i(k)}$, $i = \overline{1,s}$, $j = \overline{1, r_i}$, $k = \overline{1, l/2}$, where $f((a_{ij})_k)(b_{ij})_k = S(0, ..., (a^{(1)}_{ij})_{l/2+k} + (b_{ij})_{l/2+k}, 0, ...)$.

We got the public key $[f, (\alpha_k, \gamma_k)]$ and private key $[\beta_k, (t_{0(k)}, ..., t_{s(k)})]$, $k = \overline{1, l/2}$.

Let $x = S(0, ..., 0, x_{l/2+1}, ..., x_l)$ be the message, $x \in G_{l/2+1}$ and the public key $[f, (\alpha_k, \gamma_k)]$, $k = \overline{1, l/2}$.

Let's encrypt.

Step 6. Take a random $R = (R_1, R_2, ..., R_{l/2})$, $R_1, R_2, ..., R_{l/2l} \in \mathbb{Z}_{|F_q|}$ and calcualte

$$y_1 = \alpha'(R) \cdot x = \alpha_1'(R_1) \cdot \alpha_2'(R_2) \cdots \alpha_{l/2}'(R_{l/2}) \cdot x = S(a_1^{(1)}(R_1), a_2^{(1)}(R_2) + *, ..., a_{l/2}^{(1)}(R_{l/2}) + *,$$
$$a_{l/2+1}^{(2)}(R_1) + x_{l/2+1} + *, ..., a_l^{(2)}(R_{l/2}) + x_l + *),$$

$$y_2 = \gamma'(R) = \gamma_1'(R_1) \cdot \gamma_2'(R_2) \cdots \gamma_{l/2}'(R_{l/2}) = S(*, ..., a^{(1)}_{l/2+1}(R_1) + \beta_{l/2+1}(R_1) + *, ..., a_l^{(1)}(R_{l/2}) + \beta_l(R_{l/2}) + *).$$

The components of $(*)$ in the formula are determined in group operation of the product.

We got two vectors $(y_1, y_2)$.

Let's restore random numbers $R=(R_1,R_2,...,R_{l/2})$ to decrypt a message $x$. The parameter $a_1^{(1)}(R_1)$ is known from the $y_1$ as the first parameter and it is included in the $l/2+1$ component of $y_2$, because $a_{l/2+1}^{(1)}(R_1)=a_1^{(1)}(R_1)$.

Step 7. Calculating

$$D^{(1)}(R_1,R_2,...,R_{l/2}) = t_{0(1)} \cdot y_2 t_{s(l/2)}^{-1} = S\left(0,...,a_{l/2+1}^{(1)}(R_1)+\beta_{l/2+1}(R_1),...,a_l^{(1)}(R_{l/2})+\beta_l(R_{l/2})\right),$$

$$D^*(R) = D^{(1)}(R_1,R_2,...,R_{l/2})f(y_1) = S\left(0,...,0,\beta_{l/2+1}(R_1),a_{l/2+2}^{(1)}(R_2)+\beta_{l/2+2}(R_2)+*,...\right).$$

Let's restore $R_1$ with $\beta_{l/2+1}(R_1)$ and a use of $\beta_{l/2+1}(R_1)^{-1}$, because $\beta$ is simple. For further calculations, it is necessary to remove the components of the arrays $\alpha_1'(R_1)$ and $\gamma_1'(R_1)$ from the ciphertext $(y_1,y_2)$.

Step 8. Calculating

$$y_1^{(1)} = \alpha_1'(R_1)^{-1} \cdot y_1 = \alpha_2'(R_2) \cdot \alpha_3'(R_3) \cdots \alpha_{l/2}'(R_{l/2}) \cdot x = S\left(0,a_2^{(1)}(R_2),a_3^{(1)}(R_3)+*,...,a_{l/2}^{(1)}(R_{l/2})+*,...\right),$$

$$y_2^{(1)} = \gamma_1'(R_1)^{-1} y_2 = \gamma_2'(R_2) \cdots \gamma_{l/2}'(R_{l/2}) = S\left(*,...,a_{l/2+2}^{(1)}(R_2)+\beta_{l/2+2}(R_2)+*,...,a_l^{(1)}(R_{l/2})+\beta_l(R_{l/2})+*\right).$$

Step 9. Repeat the calculations

$$D^{(2)}(R_2,...,R_{l/2}) = t_{0(2)} \cdot y_2^{(1)} t_{s(l/2)}^{-1} = S\left(0,...,a_{l/2+2}^{(1)}(R_2)+\beta_{l/2+2}(R_2),...,a_l^{(1)}(R_{l/2})+\beta_l(R_{l/2})\right),$$

$$D^*(R) = D^{(2)}(R_2,...,R_{l/2})f(y_1^{(1)}) = S\left(0,...,0,\beta_{l/2+2}(R_2),a_{l/2+3}^{(1)}(R_3)+\beta_{l/2+3}(R_3)+*,...\right).$$

Let's restore $R_2$ with $\beta_{l/2+2}(R_2)$ and a use of $\beta_{l/2+2}(R_2)^{-1}$.

Repeating iteratively calculating after $l/2$ steps, we obtain the recovery of $R=(R_1,R_2,...,R_{l/2})$ and original message.

The correctness of such an implementation is shown in [13]. The considered encryption has several significant disadvantages. First, in the encryption algorithm, the keys $R=(R_1,R_2,...,R_{l/2})$ are loosely coupled and allow for a sequential key recovery attack. Key $R_1$ recovery through brute force attack based on brute force can be performed based on computation $\alpha_1'(R_1')$ followed by the comparison $y_1$ of the value in the coordinate $a_1^{(1)}(R_1)$ since

$$y_1 = \alpha'(R') \cdot m = S\left(a_1^{(1)}(R_1),a_2^{(1)}(R_2)+*,...,a_{l/2}^{(1)}(R_{l/2})+*,a_{l/2+1}^{(2)}(R_1)+x_{l/2+1}+*,...,a_l^{(2)}(R_{l/2})+x_l+*\right)..$$

Searching and finding $R_1'$ do not depend on the value $R_2$. Key recovery $R_2$ is possible through calculation $\alpha_2'(R_2')$ and comparison within the coordinate $a_2^{(1)}(R_2)$

$$y_1^{(1)} = \alpha_1'(R_1)^{-1} \cdot y_1 = \alpha_2'(R_2) \cdot \alpha_3'(R_3) \cdots \alpha_{l/2}'(R_{l/2}) \cdot x = S\left(0, a_2^{(1)}(R_2), a_3^{(1)}(R_3) + *, \ldots, a_{l/2}^{(1)}(R_{l/2}) + *,\right.$$
$$\left. x_{l/2+1} + *, a_{l/2+2}^{(2)}(R_2) + x_{l/2+2} + *, \ldots, a_l^{(2)}(R_{l/2}) + x_l + *\right)$$

Continuing iteratively in steps $l/2$, all keys are restored. The complexity of the attack is equal to $lq/2$. Secondly, the encryption algorithm does not use the entire scope of the generalizations of the Suzuki 2-group, but only the center $Z(G)$, which has $|Z| = q^{l/2}$ and determines the size $|x| = q^{l/2}$ of the message when encrypted.

## PROPOSED APPROACH

We propose constructing a logarithmic signature spanning the entire generalized Suzuki 2-group and implementing encryption for scoped messages across the full group. Let us examine the fundamental encryption steps.

## KEY GENERATION STAGE

We fix a large group $A_l(n, \theta) = \{S(a_1, a_2, \ldots, a_l) \mid a_i \in F_q\}$, $q = 2^n$.

Step 1. Let's build tame logarithmic signatures $\beta_k = [B_{1(k)}, \ldots, B_{s(k)}] = (b_{ij})_k = S(0, \ldots, 0, b_{ij(k)}, 0, \ldots, 0)$ of type $(r_{1(k)}, \ldots, r_{s(k)})$, $i = \overline{1, s}$, $j = \overline{1, r_{i(k)}}$, $b_{ij(k)} \in F_q$, $k = \overline{1, l}$.

Step 2. Let's set a random cover $\alpha_k = [A_{1(k)}, \ldots, A_{s(k)}] = (a_{ij})_k = S\left(a_{ij(k)}^{(1)}, a_{ij(k)}^{(2)}, \ldots, a_{ij(k)}^{(l)}\right)$ of the same type as $\beta_k$, where $a_{ij} \in A_l(n, \theta)$, $a_{ij(k)}^{(v)} \in F_q \setminus \{0\}$, $i = \overline{1, s}$, $j = \overline{1, r_{i(k)}}$, $v = \overline{1, l}$, $k = \overline{1, l}$.

Step 3. Let's generate random $t_{0(k)}, \ldots, t_{s(k)} \in A_l(n, \theta) \setminus Z$, $t_{i(k)} = S(t_{i1(k)}, \ldots, t_{il(k)})$, $t_{ij(k)} \in F^\times$, $i = \overline{0, s}$, $j = \overline{1, l}$, $k = \overline{1, l}$.

Step 4. Let's $t_{s(w)} = t_{0(w+1)}$, $w = \overline{1, l-1}$.

Step 5. Calculating $\gamma_k = [h_{1(k)}, \ldots, h_{s(k)}] = t_{(i-1)(k)}^{-1} (a_{ij})_k (b_{ij})_k t_{i(k)}$, $j = \overline{1, r_{i(k)}}$, $j = \overline{1, r_{i(k)}}$, $k = \overline{1, l}$.

We got the public key $(\alpha_k, \gamma_k)$ and private key $\left[\beta_k, (t_{0(k)}, \ldots, t_{s(k)})\right]$, $k = \overline{1, l}$.

## ENCRYPTION STAGE

Let be a message $x = S(x_1, \ldots, x_l)$ and the public key $(\alpha_k, \gamma_k)$, $k = \overline{1, l}$.

Step 6. Choose a random $R = (R_1, \ldots, R_l)$, $R_1, \ldots, R_l \in \mathbb{Z}_{|F_q|}$.

Let's set the encryption key through the mapping $R' = \pi(R_1, \ldots, R_l) = (R_1', \ldots, R_l')$.

Step 7. Calculating

$$y_1 = \alpha'(R') \cdot x = \alpha_1'(R_1') \cdot \alpha_2'(R_2') \cdots \alpha_l'(R_l') \cdot x = S\left( \sum_{k=1}^{l} \sum_{i=1, j=R'_{i(k)}}^{s(k)} a_{ij(k)}^{(1)} + x_1, \sum_{k=1}^{l} \sum_{i=1, j=R'_{i(k)}}^{s(k)} a_{ij(k)}^{(2)} + x_2 + *, ..., \right.$$

$$\left. \sum_{k=1}^{l} \sum_{i=1, j=R'_{i(k)}}^{s(k)} a_{ij(k)}^{(l)} + x_l + *, \right),$$

$$y_2 = \gamma'(R) = \gamma_1'(R_1) \cdot \gamma_2'(R_2) \cdots \gamma_l'(R_l) = S\left( \sum_{k=1}^{l} \sum_{i=1, j=R_{i(k)}}^{s(k)} a_{ij(k)}^{(1)} + \sum_{i=1, j=R_{i(1)}}^{s(1)} \beta_{ij(1)}, \sum_{k=1}^{l} \sum_{i=1, j=R_{i(k)}}^{s(k)} a_{ij(k)}^{(2)} + \right.$$

$$\left. \sum_{i=1, j=R_{i(2)}}^{s(2)} \beta_{ij(2)} + *, ..., \sum_{k=1}^{l} \sum_{i=1, j=R_{i(k)}}^{s(k)} a_{ij(k)}^{(l)} + \sum_{i=1, j=R_{i(l)}}^{s(l)} \beta_{ij(l)} + * \right),$$

$$y_3 = a'(R) = a_1'(R_1) \cdot a_2'(R_2) \cdots a_l'(R_l).$$

The components of $(*)$ in the formula are determined in the group operation of the product.

We got an output $(y_1, y_2, y_3)$.

## DECRYPTION STAGE

Let`s restore random numbers $R = (R_1, R_2, ..., R_l)$ to decrypt a message $x$.

Step 8. Calculating

$$D^{(1)}(R_1, R_2, ..., R_l) = t_{0(1)} \cdot y_2 t_{s(l)}^{-1} = S\left( \sum_{i=1, j=R_{i(1)}}^{s(1)} a_{ij(1)}^{(1)} + \beta_1(R_1), *, ..., * \right),$$

$$D^*(R) = D^{(1)}(R_1, ..., R_l) y_3^{-1} = S\left( \beta_1(R_1), a_2^{(2)}(R_2) + \beta_2(R_2) + *, ... \right).$$

Restore $R_1$ with $\beta_1(R_1)$ using $\beta_1(R_1)^{-1}$, because $\beta$ is tame. Next, it is necessary to remove the components of the array $\gamma_1'(R_1)$ from the ciphertext $y_2$.

Step 8. Calculating

$$y_2^{(1)} = \gamma_1'(R_1)^{-1} y_2 = \gamma_2'(R_2) \cdots \gamma_l'(R_l) = S\left( *, a_2^{(2)}(R_2) + \beta_2(R_2) + *, ... \right),$$

$$y_3^{(1)} = a_1'(R_1)^{-1} y_3 = a_2'(R_2) \cdots a_l'(R_l).$$

Step 9. Repeat the calculations

$$D^{(2)}(R_2, ..., R_l) = t_{0(2)} \cdot y_2^{(1)} t_{s(l)}^{-1} = S\left( 0, a_2^{(2)}(R_2) + \beta_2(R_2), *... \right),$$

$$D^*(R) = D^{(2)}(R_2, ..., R_{l/2}) \left( y_3^{(1)} \right)^{-1} = S\left( 0, \beta_2(R_2), *, ... \right).$$

Restore $R_2$ with $\beta_2(R_2)$ using $\beta_2(R_2)^{-1}$.

Step 10. Repeating iteratively calculating after $l$ steps, we get $R' = \pi(R_1, R_2, ..., R_l) = (R_1', R_2', ..., R_l')$. Thus, we obtain a recovery of the message $m = \alpha'(R_1', R_2', ..., R_l')^{-1} \cdot y_1$.

## COMPUTATIONAL VALIDATION OF PROPOSED APPROACH

Fix the generalized Suzuki group $G = A_4(n, \theta)$ over the finite field $F_q$, $q = 2^{10}$. Assume that $\theta$ is the Frobenius automorphism of $F_q, \theta: \alpha \to \alpha^2$. The group operation is defined as

$$S(a_1, a_2, a_3, a_4) S(b_1, b_2, b_3, b_4) = S(a_1 + b_1, a_2 + a_1^2 b_1 + b_2, a_3 + a_2^2 b_1 + a_1^4 b_2 + b_3, a_4 + a_3^2 b_1 + a_2^4 b_2 + a_1^8 b_3 + b_4).$$

The inverse element is determined as

$$S(a_1, a_2, a_3, a_4)^{-1} = S(a_1, a_2 + a_1^3, a_3 + a_2^2 a_1 + a_1^4 a'_2, a_4 + a_3^2 a_1 + a_2^4 a'_2 + a_1^8 a'_3)$$

where $a'_2 = a_2 + a_1^3$, $a'_3 = a_3 + a_2^2 a_1 + a_1^4 a'_2$.

Step 1. Let's construct tame logarithmic signature
$\beta_k = [B_{1(k)},...,B_{s(k)}] = (b_{ij})_k = S(0,..,0,b_{ij(k)},0,...,0)$ of type $(r_{1(k)},...,r_{s(k)})$, $i = \overline{1,s_{(k)}}$, $j = \overline{1,r_{i(k)}}$, $b_{ij(k)} \in F_q$, $k = \overline{1,l}$.

We have $l=4$ and define logarithmic signatures $\beta_k$, $k = \overline{1,4}$. It is important to choose the types $(r_{1(k)},...,r_{s(k)})$ and logarithmic signatures $\beta_k$ independently. So, let's logarithmic signatures $\beta_k$, $k = \overline{1,4}$, $s_{(k)} = 3$ have types $(r_{1(1)}, r_{2(1)}, r_{3(1)}) = (2^2, 2^2, 2^3)$, $(r_{1(2)}, r_{2(2)}, r_{3(2)}) = (2^3, 2^2, 2^2)$, $(r_{1(3)}, r_{2(3)}, r_{3(3)}) = (2^2, 2^3, 2^2)$, $(r_{1(4)}, r_{2(4)}, r_{3(4)}) = (2^2, 2^2, 2^3)$. To describe our example in detail we show the detailed representation of logarithmic signatures below.

*Table 1 Representation of logarithmic signature*

| $\beta_k = [B_{1(k)},...,B_{s(k)}] = (b_{ij})_k = S(0, b_{ij(k)}, 0, 0)$ | | | | | |
|---|---|---|---|---|---|
| **B₁₍₁₎** | $(b_{ij})_{(1)}$ | | $(b_{ij})_{(1)}$ | | $(b_{ij})_{(1)}$ |
| 0000000 | 0,0,0,0 | 0110000 | 0,α³²,0,0 | 1011100 | 0,0,α⁴⁶,0 |
| 1000000 | α⁰,0,0,0 | 1110000 | 0,α¹⁰³,0,0 | **B₃₍₃₎** | |
| 0100000 | α¹,0,0,0 | **B₂₍₂₎** | | 0000100 | 0,0,α⁴,0 |
| 1100000 | α³¹,0,0,0 | 1100000 | 0,α³¹,0,0 | 0011010 | 0,0,α¹⁷,0 |
| **B₂₍₁₎** | | 0111000 | 0,α¹⁰⁴,0,0 | 0111101 | 0,0,α²⁴,0 |
| 0100000 | α¹,0,0,0 | 0100100 | 0,α⁸,0,0 | 1001111 | 0,0,α⁹⁷,0 |
| 1010000 | α⁶²,0,0,0 | 0101100 | 0,α⁸⁵,0,0 | **B₁₍₄₎** | |
| 1101000 | α¹⁵,0,0,0 | **B₃₍₂₎** | | 0000000 | 0,0,0,0 |
| 0011000 | α³³,0,0,0 | 0010100 | 0,α⁶⁴,0,0 | 1000000 | 0,0,0,α⁰ |
| **B₃₍₁₎** | | 0010010 | 0,α⁹,0,0 | 0100000 | 0,0,0,α¹ |
| 1011000 | α⁸⁴,0,0,0 | 1001001 | 0,α³⁷,0,0 | 1100000 | 0,0,0,α³¹ |
| 1011100 | α⁴⁶,0,0,0 | 0110011 | 0,α²⁹,0,0 | **B₂₍₄₎** | |
| 0011010 | α¹⁷,0,0,0 | **B₁₍₃₎** | | 0000000 | 0,0,0,0 |
| 1110110 | α⁵⁷,0,0,0 | 0000000 | 0,0,0,0 | 0110000 | 0,0,0,α³² |
| 0011001 | α¹²³,0,0,0 | 1000000 | 0,0,α⁰,0 | 1101000 | 0,0,0,α¹⁵ |
| 0111101 | α²⁴,0,0,0 | 0100000 | 0,0,α¹,0 | 0011000 | 0,0,0,α³³ |
| 1111011 | α⁷⁵,0,0,0 | 1100000 | 0,0,α³¹,0 | **B₃₍₄₎** | |
| 0111111 | α¹¹¹,0,0,0 | **B₂₍₃₎** | | 1101000 | 0,0,0,α¹⁵ |
| **B₁₍₂₎** | | 1100000 | 0,0,α³¹,0 | 1011100 | 0,0,0,α⁴⁶ |
| 0000000 | 0,0,0,0 | 1010000 | 0,0,α⁶²,0 | 0101010 | 0,0,0,α⁸⁰ |
| 1000000 | 0,α⁰,0,0 | 1001000 | 0,0,α⁷,0 | 0111110 | 0,0,0,α⁵² |
| 0100000 | 0,α¹,0,0 | 0011000 | 0,0,α³³,0 | 0110001 | 0,0,0,α⁶¹ |
| 1100000 | 0,α³¹,0,0 | 1100100 | 0,0,α¹²¹,0 | 1110101 | 0,0,0,α⁷⁶ |
| 0010000 | 0,α²,0,0 | 1010100 | 0,0,α⁷⁹,0 | 0011011 | 0,0,0,α⁴⁰ |
| 1010000 | 0,α⁶²,0,0 | 0101100 | 0,0,α⁸⁵,0 | 0001111 | 0,0,0,α⁹⁶ |

Step 2. Construct random covers $\alpha_k = [A_{1(k)},...,A_{s(k)}] = S(a_{ij(k)}^{(1)}, a_{ij(k)}^{(2)}, a_{ij(k)}^{(3)}, a_{ij(k)}^{(4)})$, where $a_{ij} \in A_4(n,\theta)$ for the equal type as $\beta_k$, $a_{ij(k)}^{(v)} \in F_q \setminus \{0\}$, $i = \overline{1,s_{(k)}}$, $j = \overline{1,r_{i(k)}}$, $k = \overline{1,4}$.

These covers also have a field representation in a form that we have shown for the logarithmic signatures above.

Step 3. Generate random $t_{0(k)}, t_{1(k)},...,t_{s(k)} \in U(q) \setminus Z$, $s = 4$, $k = \overline{1,4}$ and $t_{s(j)} = t_{0(j+1)}$.

Step 4. Calculating the arrays $\gamma_k = [h_{1(k)},...,h_{3(k)}] = t_{(i-1)(k)}^{-1}(a_{ij})_k (b_{ij})_k t_{i(k)}$, $i = \overline{1,s_{(k)}}$, $j = \overline{1,r_{i(k)}}$, $k = \overline{1,4}$. These arrays also have a field representation in a form that we have shown for logarithmic signatures as an example above.

|Step 5. Choose a random $R = (R_1, R_2, R_3, R_4) = (20, 21, 107, 108)$, $R_k \in \mathbb{Z}_{|F_q|}$, $k = \overline{1,4}$. We obtain the following factorizations $R_k$ for each type $(r_{1(k)}, r_{2(k)}, r_{3(k)})$ in the form of $R_k = (R_{1(k)}, R_{2(k)}, R_{3(k)})$: $R_1 = (0,1,1)$, $R_2 = (5,2,0)$, $R_3 = (3,2,3)$, $R_4 = (0,3,6)$.

Step 6. Calculating $\gamma_k(R_k) = h_{1(k)}(R_{1(k)}) h_{2(k)}(R_{2(k)}) h_{3(k)}(R_{3(k)})$

$$\gamma_1(R_1) = h_{1(1)}(0) h_{2(1)}(1) h_{3(1)}(1) = S(\alpha^{107}, \alpha^{112}, \alpha^{56}, \alpha^{115})$$

$$\gamma_2(R_2) = h_{1(2)}(5) h_{2(2)}(2) h_{3(2)}(0) = S(\alpha^{46}, \alpha^{120}, \alpha^{59}, \alpha^{8})$$

$$\gamma_3(R_3) = h_{1(3)}(3) h_{2(3)}(2) h_{3(3)}(3) = S(\alpha^{20}, \alpha^{43}, \alpha^{89}, \alpha^{57})$$

$$\gamma_4(R_4) = h_{1(4)}(0) h_{2(4)}(3) h_{3(4)}(6) = S(\alpha^{87}, \alpha^{118}, \alpha^{52}, \alpha^{74})$$

*For the encryption stage, we have the following input*: a message $m \in A_4(n, \theta)$, and the public key $[f, (\alpha_k, \gamma_k)]$, $k = \overline{1,4}$. So, we are going to have ciphertext $(y_1, y_2, y_3)$ of the message $m$ as an output.

Let $m = (a^{78}, a^{10}, a^{20}, a^{21}) = S(a^{78}, a^{10}, a^{20}, a^{21})$.

Step 7. Choose a random $R = (R_1, R_2, R_3, R_4) = (20, 21, 107, 108)$.

Let's set the encryption key through the mapping

$$R' = \pi(20, 21, 107, 108) = (107, 20, 21, 108).$$

Step 8. Calculating $y_1 = \alpha'(R') \cdot m = S(\alpha^0, \alpha^{76}, \alpha^{112}, \alpha^{23})$,

$$y_2 = \gamma'(R) = \gamma_1'(R_1) \cdot \gamma_2'(R_2) \cdot \gamma_3'(R_3) \cdot \gamma_4'(R_4) = S(\alpha^{100}, \alpha^{51}, \alpha^{11}, \alpha^{52}),$$

$$y_3 = \alpha'(R) = S(\alpha^{52}, \alpha^{14}, \alpha^{86}, \alpha^{2}).$$

Output $y_1 = (\alpha^0, \alpha^{76}, \alpha^{112}, \alpha^{23})$ $y_2 = (\alpha^{100}, \alpha^{51}, \alpha^{11}, \alpha^{52})$, $y_3 = (\alpha^{52}, \alpha^{14}, \alpha^{86}, \alpha^{2})$.

We need to restore random numbers $R = (R_1, R_2, R_3, R_4)$ to decrypt a message $m$.

Step 9. Calculating

$$D^{(1)}(R_1, R_2, R_3, R_4) = t_{0(1)} y_2 t_{s(4)}^{-1} = t_{0(1)} S(\alpha^{100}, \alpha^{51}, \alpha^{11}, \alpha^{52}) t_{3(4)}^{-1} = S(\alpha^{122}, \alpha^{110}, \alpha^{26}, \alpha^{2}),$$

$$D^*(R) = D^{(1)}(R_1, ..., R_4) y_3^{-1} = S(\alpha^{34}, \alpha^{101}, \alpha^{35}, \alpha^{63}).$$

We get $\beta_1(R_1) = \alpha^{34} = (0001100)$.

Step 10. Let us restore $R_1$.

| | |
|---|---|
| 00\|01\|**100** | $R_1 = (*,*,1)$ |
| 10\|11\|100 | row 1 from $B_{3(I)}$ |
| 00\|01\|**100**−10\|11\|100=10\|**10**\|000 | $R_1 = (*,1,1)$ |
| 10\|10\|000 | row 1 from $B_{2(I)}$ |
| 10\|**10**\|000−10\|10\|000=**00**\|00\|000 | $R_1 = (0,1,1)$ |

$$R_1 = (R_{1(1)}, R_{2(1)}, R_{3(1)}) = (0,1,1) = 20$$

As a precondition for the next steps, we need to remove the components of the arrays $\gamma_1'(R_1)$, $\gamma_1'(R_2)$, $\gamma_1'(R_3)$ and $\alpha_1'(R_1)$, $\alpha_2'(R_2)$, $\alpha_3'(R_3)$ from the ciphertext $(y_2, y_3)$.

Step 11. To find $R_2$, we calculate

$$y_2^{(1)} = \gamma_1'(R_1)^{-1} y_2 = S(\alpha^{103}, \alpha^{36}, \alpha^{97}, \alpha^{113}),$$

$$y_3^{(1)} = a_1'(R_1)^{-1} y_3 = S(\alpha^{27}, \alpha^{92}, \alpha^{72}, \alpha^{38}),$$

$$D^{(2)}(R_2, R_3, R_4) = t_{0(2)} y_2^{(1)} t_{3(4)}^{-1} = S(\alpha^{27}, \alpha^{6}, \alpha^{47}, \alpha^{65}),$$

$$D^*(R) = D^{(2)}(R_2, ..., R_4)(y_3^{(1)})^{-1} = S(0, \alpha^{31}, \alpha^{20}, \alpha^{35}).$$

We get $\beta_2(R_2) = \alpha^{31} = (1100000)$.

Step 12. Let us restore $R_2$ in the same way as for $R_1$.

$$\begin{array}{ll}
110|00|\underline{00} & R_2=(*,*,0) \\
001|01|00 & \text{row 0 from } B_{3(2)} \\
110|00|\underline{00}-001|01|00=111|01|\underline{00} & R_2=(*,2,0) \\
010|01|00 & \text{row 2 from } B_{2(2)} \\
111|\underline{01}|00-010|01|00=101|\underline{01}|00 & R_2=(5,2,0) \\
\end{array}$$

$$R_2 = (R_{1(2)}, R_{2(2)}, R_{3(2)}) = (5,2,0) = 21.$$

Step 13. Let's continue calculations

$$y_2^{(2)} = \gamma_2'(R_2)^{-1} y_2^{(1)} = S(\alpha^{85}, \alpha^{29}, \alpha^{44}, \alpha^{77}),$$

$$y_3^{(2)} = a_2'(R_2)^{-1} y_3^{(1)} = S(\alpha^{113}, \alpha^{15}, \alpha^{35}, \alpha^{105}),$$

$$D^{(3)}(R_3, R_4) = t_{0(3)} y_2^{(2)} t_{3(4)}^{-1} = S(\alpha^{113}, \alpha^{15}, \alpha^{92}, \alpha^{122}),$$

$$D^*(R) = D^{(3)}(R_3, R_4)\left(y_3^{(2)}\right)^{-1} = S(0, 0, \alpha^{74}, \alpha^0).$$

We get $\beta_3(R_3) = \alpha^{74} = (1100111)$.

Step 14. Similarly, we restore $R_3$.

$$\begin{array}{ll}
11|001|\underline{11} & R_3=(*,*,3) \\
10|011|11 & \text{row 3 from } B_{3(2)} \\
11|001|\underline{11}-10|011|11=01|\underline{010}|00 & R_3=(*,2,3) \\
10|010|00 & \text{row 2 from } B_{2(2)} \\
01|\underline{010}|00-10|010|00=\underline{11}|000|00 & R_3=(3,2,3) \\
\end{array}$$

$$R_3 = (R_{1(3)}, R_{2(3)}, R_{3(3)}) = (3,2,3) = 107.$$

Step 15. Calculating $R_4$

$$y_2^{(3)} = \gamma_3'(R_3)^{-1} y_2^{(2)} = S(\alpha^{87}, \alpha^{118}, \alpha^{52}, \alpha^{74}),$$

$$y_3^{(3)} = a_3'(R_3)^{-1} y_3^{(2)} = S(\alpha^{55}, \alpha^{78}, \alpha^{110}, \alpha^{74}),$$

$$D^{(4)}(R_4) = t_{0(4)} y_2^{(3)} t_{3(4)}^{-1} = S(\alpha^{55}, \alpha^{78}, \alpha^{110}, \alpha^{121}),$$

$$D^*(R) = D^{(4)}(R_4)\left(y_3^{(3)}\right)^{-1} = S(0, 0, 0, \alpha^{36}).$$

We get $\beta_4(R_4) = \alpha^{36} = (0000011)$.

Step 16. Similarly, we restore $R_4$.

$$\begin{array}{ll}
00|00|\underline{011} & R_4=(*,*,6) \\
00|11|011 & \text{row 6 from } B_{3(1)} \\
00|00|\underline{011}-00|11|011=00|\underline{11}|000 & R_4=(*,3,6) \\
00|11|000 & \text{row 3 from } B_{2(1)} \\
00|\underline{11}|000-00|11|000=\underline{00}|00|000 & R_4=(0,3,6) \\
\end{array}$$

$$R_4 = (R_{1(4)}, R_{2(4)}, R_{3(4)}) = (0,3,6) = 108.$$

Step 17. Permute $R' = \pi(20, 21, 107, 108) = (107, 20, 21, 108)$ and restore $m = \alpha'(R')^{-1} \cdot y_1 = S(a^{78}, a^{10}, a^{20}, a^{21})$.

# SECURITY ANALYSIS

We examine the following attack vectors. At this stage of research, our objective is not to provide a comprehensive security analysis, but rather to evaluate potential brute force attacks and quantify their computational complexity.

Attack 1. Brute force attack on the ciphertext.

We try to decipher the text $y_1' = \alpha'(R') \cdot m$ by selecting $R = (R_1, ..., R_l)$. The covers are chosen at random and the value is determined by multiplication in a group with no coordinate restrictions. The complexity of the attack is $q^l$ if our attack model implies a known text.

Attack 2. Brute force attack on the ciphertext $y_2$.

To match $y_2$ we need to select $R = (R_1, ..., R_l)$ respectively. The calculations over the vectors $\alpha_1'(R_1), ..., \alpha_l'(R_l)$ define the values of the coordinates $y_2$. The keys $R_1, ..., R_l$ are bound. Any changes in their consistency lead to change $y_2$. In this case the attack on the key $R$ has a same complexity equal to $q^l$.

Attack 3. Brute force attack on the ciphertext $y_3$.

To match $y_3$ we need to select $R = (R_1, R_2, R_3)$ respectively. The values of the coordinates $y_3$ are defined by calculations over the vectors $\alpha_1'(R_1), ..., \alpha_l'(R_l)$. The keys $R_1, ..., R_l$ are bound. Any changes in their consistency lead to change $y_3$. In this case the attack on the key $R$ also has a same complexity equal to $q^l$.

Attack 4. Brute force attack on the $(t_{0(k)}, ..., t_{s(k)})$.

It's quite clear that the brute force attack on $(t_{0(k)}, ..., t_{s(k)})$ is a general for the MST family. An optimistic complexity estimation for the calculation in the field $F_q$ over the group center $Z(G)$ is equal to $q$. Our proposal uses a calculation in the whole group $|G| = q^l$. This improvement increases the complexity of the attack on $(t_{0(k)}, ..., t_{s(k)})$ to $q^l$.

# CONCLUSION

Our approach is to use the generalized Suzuki 2 group as a platform for MST3 cryptosystem. The full group encryption $G = A_l(n,\theta)$ with associated keys $R = (R_1,...,R_l)$ improves security parameters and attack complexity to $q^l$. Our improvement also extends the logarithmic signature to the entire generalized Suzuki 2 group $A_l(n,\theta)$ $|A_l(n,\theta)| = q^l$. We also bind the keys and change the encryption algorithm. This improvement provides additional security against sequential recovery attacks.